\documentclass[12pt]{article}

\usepackage{epsf}
\usepackage{latexsym,euscript}
\usepackage[dvips]{graphicx}
\usepackage{amsmath}
\usepackage{amsfonts}
\usepackage{amssymb, epsfig}

\DeclareMathOperator{\arcsinh}{arcsinh}

\usepackage{color,soul}

\begin{document}

\begin{center}
{\Large \textbf{The large $N$ limit of $SU(N)$ integrals in lattice models}}

\vspace*{0.6cm}
\textbf{O.~Borisenko${}^{\rm a}$\footnote{email: oleg@bitp.kiev.ua},
V.~Chelnokov${}^{\rm a}$\footnote{email: chelnokov@bitp.kiev.ua},
S.~Voloshyn${}^{\rm a}$\footnote{email: billy.sunburn@gmail.com}}

\vspace*{0.3cm}
{\large \textit{${}^{\rm a}$ Bogolyubov Institute for Theoretical
Physics, National Academy of Sciences of Ukraine, 03143 Kyiv, Ukraine}}
\end{center}

\begin{abstract}
The standard $U(N)$ and $SU(N)$ integrals are calculated in the large $N$ limit. 
Our main finding is that for an important class of integrals this limit is different 
for two groups. We describe the critical behaviour of $SU(N)$ models and discuss implications of 
our results for the large $N$ behaviour of $SU(N)$ lattice gauge theories at finite 
temperatures and non-zero baryon chemical potential. The key ingredients of our approach are 
1) expansion of the integrals into a sum over irreducible representations and 
2) calculation of sums over partitions of $r$ of products of dimensions 
of two different representations of a symmetric group $S_r$. 
\end{abstract}

\section{Introduction}

Let $U \in G$, where $G=U(N), SU(N)$ and $A,B$ are two arbitrary $N\times N$ matrices. 
$\mbox{Tr} U$ will denote the character of the fundamental representation of
$G$ and $dU$ - the normalized (reduced) Haar measure on $G$.
In this paper we evaluate the following integral over $G$ in the large $N$ limit 
for some particular choices of $A$ and $B$
\begin{equation}
Z_G(A,B) \ = \ \int_G dU \ \exp \left [ \frac{N h}{2} \left ( {\rm Tr} A U 
+ {\rm Tr} B U^{\dagger}  \right )   \right  ] \ . 
\label{Gint_def}
\end{equation}
Though the constant $h/2$ can be incorporated into $A$ and $B$ it is convenient to keep it in the present form. 
Many cases of the integral (\ref{Gint_def}) are well-known in the context of the lattice gauge theory (LGT)
and classical spin models. Let us briefly summarize the known facts. 
When $A=B=I$ the integral in (\ref{Gint_def}) gives the exact solution of two-dimensional LGT with the Wilson action 
in the thermodynamical limit. 
The large $N$ solution of this integral had been obtained in \cite{gross_witten,wadia} and is known as the Gross-Witten-Wadia (GWW) solution. 
The third order phase transition (often referred to as the GWW transition) was found in this limit. 
More general case with $B=A^{\dagger}$ in the large $N$ limit was solved in \cite{brezin_gross}. 
The actual calculations in \cite{brezin_gross} were performed for $U(N)$ group. This solution also revealed the existence of the third 
order phase transition. Integrals of type (\ref{Gint_def}) can be considered as a generating functional for many group integrals 
of polynomial type \cite{creutz,carlsson08}. They also appear in the mean-field treatment of LGT 
and principal chiral models \cite{lgt_mean_field,pcm_strong_coupl,pcm_mean_field}. Important usage of such integrals can be found 
in many effective models describing the Polyakov loop interactions in finite temperature LGT. For example, 
if one takes $A=e^{\mu} I$ and $B=e^{-\mu}I$ then the expression in the exponent of the integrand is nothing but the leading 
term in the expansion of the quark determinant at large masses, where ${\rm Tr}U$ can be interpreted as the Polyakov loop. 
Such broad applications make it necessary to have a deep 
understanding of the properties of the integral in different regimes. Particularly important is the knowledge of the integral 
in the large $N$ limit and, indeed there exists a huge literature devoted to this limit when $G=U(N)$ 
(see, {\it e.g.} Ref.\cite{dunne} and references therein). $SU(N)$ integrals are also well studied for finite $N$ and, 
in some cases also in the large $N$ limit \cite{pcm_mean_field,dunne,rossi81,rossi84,lucini_panero, rossi96_rev}. 
Still, we think the large $N$ limit of $SU(N)$ integrals remains largely unexplored. This is probably due to the fact that since 
in the large $N$ limit $SU(N)$ reduces to $U(N)$, the standard claim is that the same holds for integrals (\ref{Gint_def}). 
The purpose of this paper is to show that it is not always the case. Moreover, based on examples we consider below 
we can conclude that the large $N$ limit of integrals of type (\ref{Gint_def}) is the same for $U(N)$ and $SU(N)$ only 
in exceptional cases. Probably, there is only one such case when integrals are equal for all values of $h$, namely ${\rm Re}A={\rm Re}B$. 

There are several methods developed to compute the large $N$ limit of group integrals \cite{gross_witten,wadia,goldschmidt80,rossi82,carlson84}. 
We shall not discuss these methods because here we use a different approach based on the expansion of the group integrals in a sum over 
dimensions $d(\lambda)$ of representations of a permutation group $S_r$. Precisely, the integral in (\ref{Gint_def}) 
can be evaluated for $G=SU(N)$ as \cite{weingarten_sun}
\begin{align}
Z_{\mathrm{SU}(N)}(A,B) &=  \sum_{r=0}^{\infty} \sum_{q=-\infty}^{\infty} 
\ \left ( \frac{N h}{2} \right )^{2r+N|q|} \nonumber \\
&\phantom{=} \times
\sum_{\lambda \vdash r} \ \frac{d(\lambda)d(\lambda + |q|^N)}{r!(r+N|q|)!} 
\ \frac{s_{\lambda}(A B)}{s_{\lambda}(1^N)} 
\left ( \mbox{det} C \right )^{|q|} , 
\label{sun_1link_res}
\end{align}
where $C=A$ if $q>0$ and $C=B$ if $q<0$. $s_{\lambda}$ is the Schur function defined on the eigenvalues of $AB$. 
Summation goes over all partitions $\lambda$ of $r$: $\lambda=(\lambda_1,\lambda_2,\cdots,\lambda_N)$ is a partition $\lambda\vdash\ r$, 
$\lambda_1\geq\lambda_2\geq \cdots \geq \lambda_N\geq 0$ and $\sum_{i=1}^{l(\lambda)} \lambda_i \equiv |\lambda|= r$, 
where $l(\lambda)$ is the length of the partition $\lambda$. 
We have used the short-hand notation $\lambda+q^N=(\lambda_1+q,\cdots,\lambda_N+q)$. 
When $G=U(N)$ one should take the only term with $q=0$ in expression (\ref{sun_1link_res}). 
It can be proven that the resulting formula agrees with the $U(N)$ result \cite{wettig02,novak20}. Furthermore, the corresponding 
representation for $U(N)$ integral has recently been used to rigorously prove the large $N$ asymptotic expansion 
in the strong coupling region (small $h$ region) \cite{novak20}.
Our method relies on calculating large $N$ limit of sums over all partitions $\lambda$. 
This is particularly simple for examples we consider in this paper when the sum over $\lambda$ can be reduced to 
the calculation of the following expression 
\begin{equation} 
Q_N(r,q) \ = \  \sum_{\lambda \vdash r} \ d(\lambda) \ d(\lambda + |q|^N) \ . 
\label{QSUN}
\end{equation}

This paper is organized as follows. In Sec.~2 we consider the simplest case when one of the matrices $A$ or $B$ vanishes. 
Sec.~3 treats more general case $A=e^{\mu} C$, $B=e^{-\mu}C^{-1}$ for real and imaginary $\mu$. 
Results and applications are discussed in Sec.~4. 
In the Appendix we study large $N$ expansion of $Q_N(r,q)$ defined in Eq.(\ref{QSUN}) and related functions.

\section{Reduced integral: $B=0$}

In this Section we explore the simplest case with $B=0$ and an arbitrary matrix $A$ 
\begin{eqnarray}
Z_G\left ( h \right )  \ = \
\int_G  dU  \ e^{  \frac{N h}{2} \ {\rm {Tr}} A U } \ , 
\label{Z_def_sun}
\end{eqnarray}
which can be solved exactly and which shows that the large $N$ limit differs for two groups. Indeed, if $G=U(N)$ 
one gets the trivial result for all $N$, namely $Z_{U(N)}(h)=1$. 
Instead, if $G=SU(N)$ we obtain from (\ref{sun_1link_res})
\begin{eqnarray}
Z_{SU(N)}\left ( h \right ) =  \sum_{q=0}^{\infty} \ \frac{\left ( \frac{N h}{2} \right )^{qN}}{(q N)!} \ 
\left ( \mbox{det} A \right )^{q} \ Q_N(0,q) \ .  
\label{Z_sun_res}
\end{eqnarray}
Keeping notation $h$ for $h(\prod_{i=1}^N a_i)^{\frac{1}{N}}$ with $a_i$ - eigenvalues of $A$ and using (\ref{QSUN_r0}) 
the last equation is presented as 
\begin{eqnarray}
Z_{SU(N)}\left ( h \right ) =  \sum_{q=0}^{\infty} \ \left ( \frac{N h}{2} \right )^{qN} \ A_N(q) \ , 
\label{Z_sun_B0}
\end{eqnarray}
where the function $A_N(q)$ is defined in (\ref{ANq}). With help of Eq.(\ref{ANq_as}), one gets 
\begin{equation}
Z_{SU(N)}\left ( h \right )  \ = \ \sum_{q=0}^{\infty} \ \exp \left [ N^2 \left ( x\ \ln\frac{h}{2} + f(x) \right ) \right ] \ , 
\label{Z_sun_limit}
\end{equation}
where $f(x)$ is given by Eq.(\ref{fx_as}) and $x=q/N$. In the $N\to\infty$ limit the summation over $q$ can be replaced 
by the integration and we obtain for the free energy 
\begin{equation}
F\left ( h \right )  \ = \ \lim_{N\to\infty} \frac{1}{N^2} \ \ln Z_{SU(N)}\left ( h \right )  \ = \ x_0 \ \ln\frac{h}{2} + f(x_0)  \ , 
\label{F_sun_limit}
\end{equation}
where $x_0$ is determined from the saddle-point equation 
\begin{equation}
1+\ln \frac{h}{2} + x_0 \ln x_0 - (1+x_0)\ln (1+x_0) \ = \ 0 \ .  
\label{saddle_eq_sun}
\end{equation}
There is no real solution to this equation if $h<2/e$. Therefore, $F(h<2/e)=0$. 
When $h=2/e$ the real solution $x_0=0$ appears and solution becomes non-trivial in the region $h>2/e$. 
Approximate solution in the vicinity of $h = 2/e +0^+$ and the free energy can be written as 
\begin{eqnarray}
x_0  \approx - y + \frac{1}{2\left ( 1+W_{-1}\left ( -z/e \right )\right )} \ y^2  
+ \frac{4W_{-1}\left ( -z/e \right )\left ( W_{-1}\left ( -z/e \right )-1\right )-5}{24 
\left ( 1+W_{-1}\left ( -z/e  \right ) \right )^3} \ y^3   \ , 
\label{x0_solut}
\end{eqnarray}
\begin{equation}
F(h) = - \frac{1}{4} \ \left ( 1+ 2 W_{-1}\left ( -z/e \right ) \right ) y^2 
+ \frac{1}{6} \ y^3 + {\cal O}\left ( y^4 \right )   \ , 
\label{Fr_reduced}
\end{equation}
where $y=\frac{z}{W_{-1}(-z/e)}$, $z=1+\ln\frac{h}{2}$ and $W_{-1}(-z/e)$ is the lower branch of the Lambert function. 

\begin{figure}[t]
\centerline{{\epsfxsize=6.6cm \epsfbox{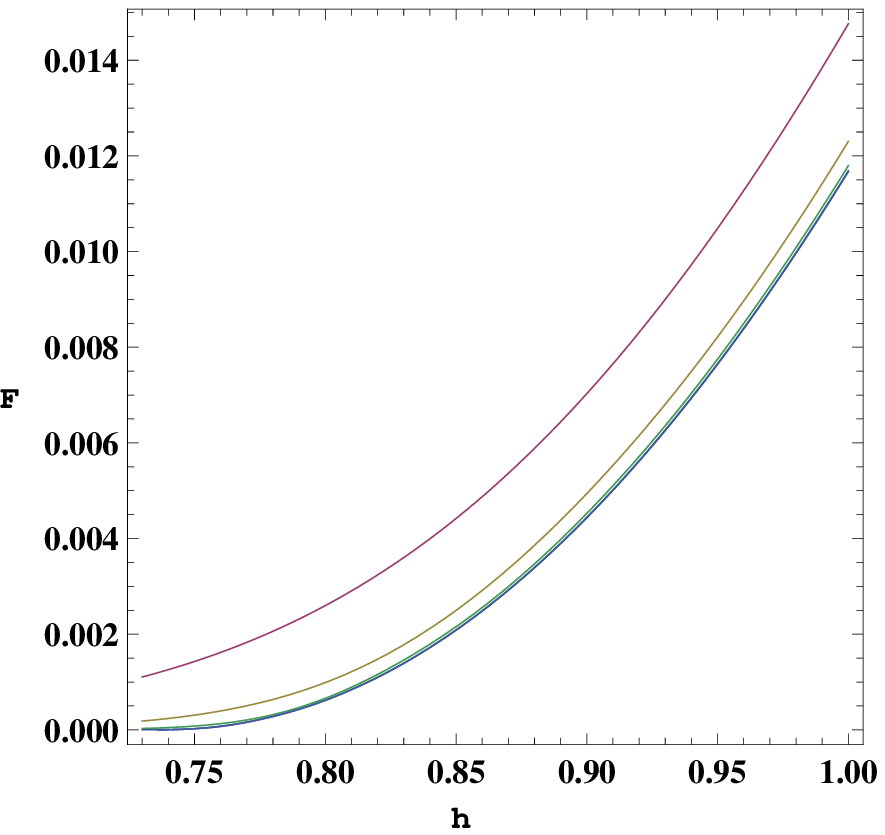}} {\epsfxsize=6.25cm\epsfbox{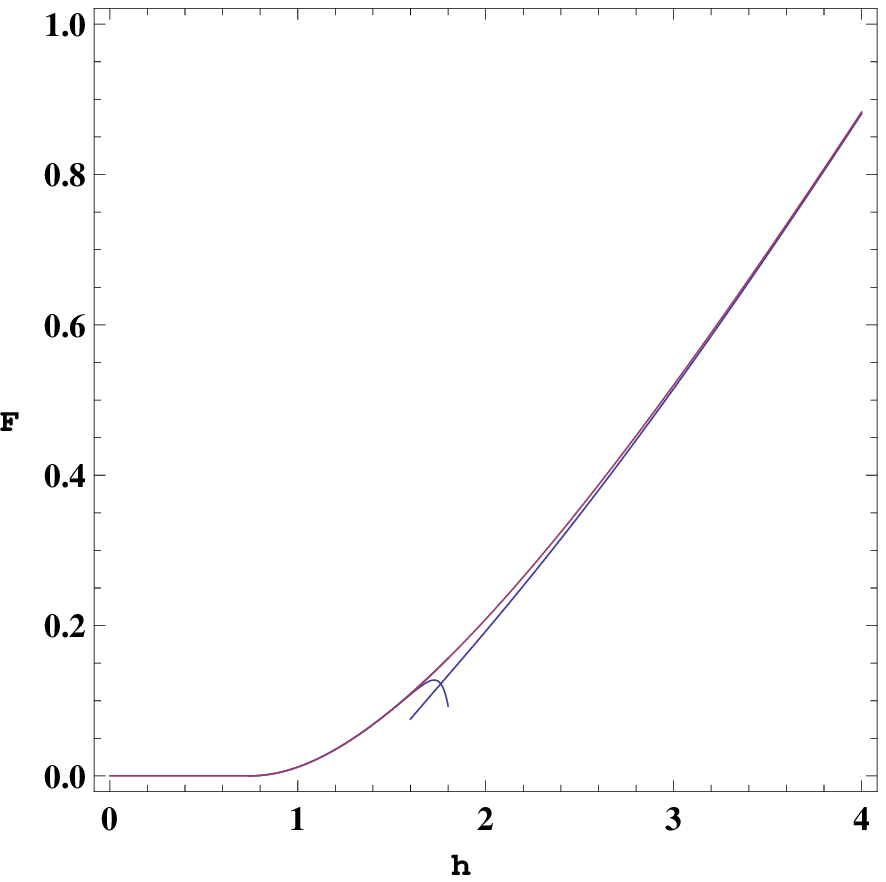}}}
\caption{\label{plot12} Comparision of the analytical solution and numerical computations. Left panel: 
Blue line presents analytical solution near critical point. Numerical computations 
are performed for $N=10$ (red line), $N=20$ (yellow line), $N=40$ (green line). $N=100$ data are practically indistinguishable from 
the blue analytical line. Right panel: Blue lines present analytical solution (\ref{Fr_reduced}) (lower branch) and (\ref{Fren_large_h}) 
(upper branch). Numerical computation is performed for $N=40$ (red line)}
\end{figure}

Near critical point $z=0$ the Lambert function can be expanded as $W_{-1}(-z/e)\approx \ln z/e - \ln(-\ln z/e)+\cdots$. 
It follows that the free energy and its first two derivatives vanish in the critical point while the third derivative 
diverges like $1/(z\ln^2 z)$. The model exhibits the third order phase transition from $U(N)$ regime with vanishing free 
energy to a regime with a non-trivial dependence on $h$ which we term $SU(N)$ regime in what follows. On the left panel of Fig.\ref{plot12} 
we compare analytical solution (\ref{Fr_reduced}) with numerical computations performed with Eq.(\ref{Z_sun_B0}) 
for various values of $N$ near critical point.

When $h$ is large the saddle-point solution and the corresponding free energy read 
\begin{align}
\label{x0_large_h}
x_0  &=  \frac{h}{2} - \frac{1}{2} - \frac{1}{4h} + {\cal O}\left ( h^{-2}  \right )  \ , \\  
\label{Fren_large_h}
F(h) &=  \frac{h}{2} - \frac{1}{4} \left ( 2 \ln\frac{h}{2} + 3  \right ) - \frac{1}{12h}  + {\cal O}\left ( h^{-2}  \right )  \ . 
\end{align}
We could not find an exact solution of the saddle-point equation for all values of $h$. Nevertheless, we think the phase transition 
at $h=2/e$ is the  only phase transition occurring in the large $N$ limit. To see this we have performed numerical computations 
of the partition function (\ref{Z_sun_res}) and its first three derivatives for $h>1.4$ and found no evidence on non-analyticity 
of the free energy in this regime. On the right panel of Fig.\ref{plot12} we compare analytical solutions for the free energy, 
Eqs.(\ref{Fr_reduced}) and (\ref{Fren_large_h}), with numerical computations performed with Eq.(\ref{Z_sun_B0}) for $N=40$. 

Another possible approach is to express the constant $h$ as a function of $x_0$ from (\ref{saddle_eq_sun}). 
Then, the free energy can be studied as a function of $x_0$. This study also reveals that the point $x_0=0$ 
is the only critical point of the free energy.

\section{Integral with $A=e^{\mu} C$, $B=e^{-\mu} C^{-1}$}

The next example we explore deals with the integral 
\begin{equation}
Z_G\left ( h,\mu \right )  \ = \ \int_G  dU  \ e^{  \frac{N h}{2} \ ( e^{\mu} {\rm {Tr}}C U  + 
 e^{-\mu} {\rm {Tr}}C^{-1} U^{\dagger} ) } \ . 
\label{Z_quark}
\end{equation}
The example studied in the previous section with $A=I$
is $\mu\to\infty$ limit of the integral (\ref{Z_quark}) such that $h e^{\mu}={\rm const}$. 
The integral (\ref{Z_quark}) arises in the strong coupling and heavy quark limit of $SU(N)$ LGT 
at finite baryon density. Parameter $h$ can be related to the quark mass and ${\rm {Tr}}U$ is the traced Polyakov loop. 
Hence, the model (\ref{Z_def_sun}) is a heavy dense limit of QCD. Keeping notation $e^{\mu}$ for $e^{\mu}(\prod_{i=1}^N c_i)^{\frac{1}{N}}$ 
with $c_i$ - eigenvalues of $C$ and using (\ref{sun_1link_res}) and (\ref{QSUN}) the result 
of the integration in (\ref{Z_quark}) is presented in the form 
\begin{equation}
Z_{SU(N)}\left ( h,\mu \right ) \ = \ \sum_{q=-\infty}^{\infty} \ e^{N \mu q}  \sum_{r=0}^{\infty} \ 
\frac{ (\frac{N h}{2})^{2r+|q|N}}{r!(r+|q|N)!} \ Q_N(r,|q|)  \ . 
\label{SUN_integral}
\end{equation}
If $G=U(N)$ only term with $q=0$ contribute and all dependence of the integral on $\mu$ drops. 
The integral reduces to $2d$ $U(N)$ LGT solved in \cite{gross_witten,wadia}. 
Using representation (\ref{QSUN_small}) we rewrite $SU(N)$ integral (\ref{SUN_integral}) as 
\begin{equation}
Z_{SU(N)}\left ( h,\mu \right ) \ = \ \sum_{q=-\infty}^{\infty} \ \left ( \frac{N h}{2} \right )^{|q|N} \ 
e^{N \mu q} \ A_N(|q|) \  H_N(h,|q|) \ , 
\label{SUN_integral_1}
\end{equation}
where we denoted 
\begin{equation}
H_N(h,|q|) \ = \ \sum_{r=0}^{\infty} \ \frac{ \left ( \frac{N h}{2} \right )^{2r}}{r!} \ B_N(r,|q|)  \ .
\label{Hfunct1}
\end{equation}
The function $H_N(h,|q|)$ is studied in the Appendix. Next, we proceed as in previous section. Combining 
Eqs.(\ref{Hfunct_res},\ref{ANq_as},\ref{fx_as}) we write down the partition function in the form 
\begin{eqnarray}
\label{SUN_integral_2}
Z_{SU(N)}\left ( h,\mu \right ) = \sum_{q=-\infty}^{\infty} \ \exp \left [ N^2 F(h,\mu,x) \right ] \ , \
x=\frac{q}{N} \ ,  \\ 
F(h,\mu,x) =  \frac{h^2}{4}  + \left(\frac{3}{2} + \mu + \ln \frac{h}{2} \right) x
+ \frac{1}{2} x^2\ln x - \frac{1}{2} (1+x)^2\ln (1+x) - P(x) \ ,
\label{Fren_sun1}
\end{eqnarray}
where $P(x)$ is given in Eq.(\ref{Px_def}).
Replacing the summation by the integration one obtains in the large $N$ limit 
\begin{equation}
F(h,\mu) \ = \ F(h,\mu,x_0) \ . 
\label{Fren_sun2}
\end{equation}
The saddle-point equation is given by (we suppose $\mu \geq 0$)
\begin{equation}
1+\mu +\ln \frac{h}{2} + x_0 \ln x_0 - (1+x_0)\ln (1+x_0) - \frac{\partial P(x_0)}{\partial x_0} \ = \ 0 \ .  
\label{saddle_eq_sun1}
\end{equation}
If $h\leq 1$ and $\mu$ is sufficiently small there is no solution to this equation and the free energy 
equals to the $U(N)$ free energy, namely $F(h,\mu)=h^2/4$. When $\mu$ grows a non-trivial solution appears 
which provides maximum of the integrand in (\ref{SUN_integral_2}). 
This solution can be constructed using expansion (\ref{x_small_series}) for the function $P(x)$. 
With notations 
\begin{equation}
y \ = \ \frac{z}{W_{-1} \left ( -\frac{z}{e \sqrt{1 - h^2}} \right ) } \  \ , \   \ 
z \ = \ \mu - \ln \frac{1 + \sqrt{ 1 - h^2}}{h} + \sqrt{1 - h^2} 
\label{notation_sun1}
\end{equation}
the solution and the free energy can be written as
\begin{eqnarray}
x_0  =  -y + \frac{1}{2\left ( 1 - h^2 \right )^{3/2} \left  (1+ W_{-1} \left ( - \frac{z}{e \sqrt{1 - h^2}} \right ) \right ) } \ y^2 
+ {\cal O}\left ( y^3 \right ) \ , 
\label{x0_solut_small_h}
\end{eqnarray}
\begin{eqnarray}
F(h,\mu) =  \frac{h^2}{4} -  
\frac{1}{4} \  \left ( 1 + 2 W_{-1} \left ( - \frac{z}{e \sqrt{1- h^2}} \right ) \right ) \ y^2  -   
\frac{1}{6(1 - h^2)^{3/2}} \ y^3 + {\cal O}\left ( y^4 \right ) . 
\label{Fren_sun_small_h}
\end{eqnarray}
One finds a third order phase transition along the critical line which is given by the equation 
\begin{equation}
z \ = \ 0 \  \  \Rightarrow \  \mu \ = \  \ln \frac{1 + \sqrt{ 1 - h^2}}{h} - \sqrt{1 - h^2} \ . 
\label{critical_line}
\end{equation}
When $\mu = 0$, the critical value of $h$ equals $h_c=1$ and coincides with $U(N)$ critical point. 
In the region $h>1$ the solution is found with the help of the expansion (\ref{x_large_series}) for the function $P(x)$. 
The solution and the corresponding free energy appear to be 
\begin{eqnarray}
\label{x0_solut_large_h}
&& x_0 =   h \sinh \mu  - \tanh \mu \left( \frac{1}{2} -   \frac{\cosh 2 \mu - 5   }{48 h( \cosh \mu )^3}  -   
   \frac{\cosh 2 \mu - 2 }{24 h^2(\cosh \mu )^6}  \right.   \\
&& \left. - \frac{  3 \cosh 6 \mu  - 902 \cosh 4 \mu + 14845\cosh 2 \mu  -18810  }{184320 h^3(\cosh \mu )^9} \right )
 + {\cal O} \left(\frac{1}{h^4}\right)  \ ,  \nonumber 
\end{eqnarray}
\begin{eqnarray}
&&F(h,\mu) =  h \cosh \mu  -  \frac{1}{2} \left( \frac{3}{2}+ \ln h \cosh \mu \right) -  
\frac{(\tanh \mu)^2}{24 h} \left( \frac{1 }{\cosh \mu}   + \frac{1}{2 h(\cosh \mu)^4} \right.  \nonumber \\
&& \left. +  \frac{\cosh4 \mu - 176 \cosh 2 \mu    + 783 }{1920 h^2 ( \cosh \mu)^7}+ \frac{3 \cosh 4 \mu  - 
74 \cosh 2 \mu + 143   }{320  h^3 (\cosh \mu)^{10}} \right) +  {\cal O} \left(\frac{1}{h^4}\right).  
\label{Fren_sun_large_h}
\end{eqnarray}
This result can be checked with the help of the asymptotic expansion (\ref{detsun_as}) valid at large values of $h$. 
Using Eq.(\ref{detsun_as}) one can reproduce leading three terms in Eq.(\ref{Fren_sun_large_h}).

\begin{figure}[t]
\centerline{{\epsfxsize=8cm \epsfbox{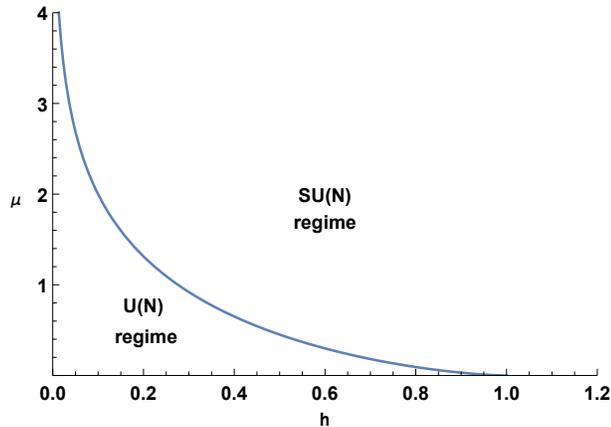}}}
\caption{\label{crit_line} Critical line of the third order phase transition from $U(N)$ regime to $SU(N)$ one. 
In $U(N)$ regime the free energy does not depend on $\mu$. In the limit $\mu\to\infty$ the critical curve 
is well approximated by the formula $\mu\approx \ln\frac{2}{h} - 1$.}
\end{figure}

Let us add a few comments: 
\begin{itemize}
\item 
When $\mu=0$ the solution (\ref{x0_solut_large_h}) vanishes, the free energy (\ref{Fren_sun_large_h}) reduces 
to the $U(N)$ free energy and equals the GWW solution. 

\item 
Though we could not prove it analytically we think the transition from the regime described by Eq.(\ref{Fren_sun_small_h}) 
to the large $h$ regime in Eq.(\ref{Fren_sun_large_h}) is smooth and is not accompanied by any phase transition. We have 
numerical evidence which supports this statement. If so, the large $N$ limit exhibits one third order phase transition 
from $U(N)$ to $SU(N)$ regime and whose critical line is given by Eq.(\ref{critical_line}). This line is shown in Fig.~\ref{crit_line}. 

\item 
Another approach is to express the parameter $\mu$ as a function of $x$ from the saddle-point equation (\ref{saddle_eq_sun1}). 
Then, one can study the free energy in the $(h,x)$-plane. This study also reveals that there is only one phase transition 
described by the critical line (\ref{critical_line}) above which the free energy is an analytical function. 

\item 
In the case of an imaginary $\mu$ one can recover again the GWW $U(N)$ solution. Indeed, when $\mu$ is imaginary, there is no 
solution for the saddle-point equation on the real line. Therefore, the maximum of the exponent in Eq.(\ref{SUN_integral_2}) 
is reached for $x=0$ and this recovers the $U(N)$ case. We have also performed numerical computations of the free energy 
for several values of imaginary $\mu$ and various values of $N\in[10-35]$. These computations clearly demonstrate that the free 
energy converges to the $U(N)$ free energy with $N$ growing. 

\item 
In the heavy dense limit $\mu\to\infty, h\to 0$ such that $h e^{\mu}={\rm const}$ one obtains from Eq.(\ref{critical_line}) 
that the critical point is given by $he^{\mu}=2/e$. This agrees with the result of the previous Section. 

\end{itemize}

\section{Summary}

In this paper we have presented calculations of some classical $SU(N)$ integrals in the large $N$ limit. 
Our most essential result is that in many important cases this limit differs from the conventional limit obtained 
by calculating the same integrals over $U(N)$. The important findings and open issues can be shortly summarized as follows:  

\begin{itemize} 

\item 
In all cases studied we have found the third order phase transition from the $U(N)$ regime to the $SU(N)$ regime. 
Unlike the GWW phase transition, the third derivative of the free energy is infinite on the critical line. 

\item 
It would be important to solve the integral (\ref{Gint_def}) in the large $N$ limit for arbitrary matrices $A$ and $B$. 
Based on examples studied here we {\it conjecture} that this limit differs for two groups unless ${\rm Re}A={\rm Re}B$. 

\item 
Our main tool used to get the large $N$ solution for $SU(N)$ integrals relies on 1) the representation (\ref{sun_1link_res}) 
which is exact and 2) calculation of the function $Q_N(r,q)$ (\ref{QSUN}). We have been able to calculate this function 
for all $N$ and $q$ and for $r=1,2,\ldots,11$. This is sufficient to establish the existence of the phase transition 
and calculate the critical line. 

\item 
We have also {\it conjectured} an exact form of the function $Q_N(r,q)$ at large $N$ given by Eq.(\ref{QN_exact_largeN}). 
It remains a challenge to prove this form rigorously. 

\item 
We have derived leading terms in an asymptotic expansion for the $SU(N)$ integral at large $h$, Eq.(\ref{detsun_as}). 
This expansion has been used as an independent check of the free energy expansion given by Eq.(\ref{Fren_sun_large_h}).

\end{itemize}

Let us also emphasize the approach used here allows us to reproduce the GWW $U(N)$ solution in a very effective and simple manner. 
Finishing this paper we would like to briefly outline the perspectives for future work.  
As follows from the results of Sec.~3, one could expect the different large $N$ limits for $U(N)$ and $SU(N)$ groups 
whenever the symmetry between $U$ and $U^{\dagger}$ is broken (in physical terms it means broken charge conjugation symmetry). 
This is a typical situation in QCD at finite baryon chemical potential. For example, charge conjugation symmetry is broken in 
all effective Polyakov loop models computed in lattice QCD at finite density. Consider, as the simplest and well-known example, 
the following effective action 
\begin{eqnarray} 
\label{PL_action}
S \ = \  \beta_{eff} \ \sum_{x,\nu} {\rm Re}{\rm Tr}U(x){\rm Tr}U^{\dagger}(x+e_{\nu}) + 
N_f h \sum_x \left ( e^{\mu} {\rm {Tr}}U(x) +  e^{-\mu} {\rm {Tr}}U^{\dagger}(x) \right ) 
\end{eqnarray}
which describes the interaction between Polyakov loops ${\rm {Tr}}U(x)$ at finite temperature in the strong coupling region 
and in the presence of $N_f$ heavy degenerate quark flavours. This and other similar effective actions have been a subject 
of numerous analytical and numerical investigations. In particular, the large $N$ limit has been studied in Ref.\cite{christensen12} 
using the mean-field approach which is believed to be exact in this limit due to the factorization property. The resulting mean-field 
integrals have been computed over $U(N)$ group. As we have seen, however when $\mu$ is non-vanishing the $SU(N)$ integrals 
differ from $U(N)$ ones. Therefore, the analysis of \cite{christensen12} should be re-examined for the $SU(N)$ model. 
The same presumably refers to Polyakov loop models with exact static quark determinant. 

In Refs.\cite{un_dual,pldual} we have derived the dual representation for the Polyakov loop models of the type (\ref{PL_action})
for all $N$. The function $Q_N(r,q)$ plays a central role in the dual construction and appears as a result of the group integration. 
Combining the dual representation of \cite{un_dual,pldual} with the large $N$ representation of $Q_N(r,q)$ derived here one can construct 
the large $N$ limit of the model (\ref{PL_action}) and similar ones. These calculations will be reported elsewhere.

\section{Appendix: function $Q_N(r,q)$}

The function $Q_N(r,q)$ plays a central role in our method of calculation of the large $N$ limit. 
It appears as a result of the evaluation of the following group integral 
\begin{equation}
Q_N(r,q) = \int_G dU \ ({\rm Tr} U)^r \ ({\rm Tr} U^{\dagger})^{r+qN} = 
\int_G dU \ ({\rm Tr} U)^{r+qN} \ ({\rm Tr} U^{\dagger})^r \ . 
\label{Q_def_app}
\end{equation}
The result of the integration over $SU(N)$ reads \cite{weingarten_sun}
\begin{equation} 
Q_N(r,q) \ = \  \sum_{\lambda \vdash r} \ d(\lambda) \ d(\lambda + |q|^N) \ , 
\label{QSUN_A}
\end{equation}
where $\lambda+q^N=(\lambda_1+q,\cdots,\lambda_N+q)$, all notations are explained after Eq.(\ref{sun_1link_res}) 
and the dimension of the $\lambda$ representation of the symmetric group $S_r$ is given by 
\begin{equation}
d(\lambda) \ = \ r! \
\frac{\prod_{1\leq i<j \leq l(\lambda)}(\lambda_i-\lambda_j+j-i)}
{\prod_{i=1}^{l(\lambda)}(\lambda_i+l(\lambda)-i)!} \ . 
\label{dim_l}
\end{equation}
For $U(N)$ the integral is only non-vanishing if $q=0$. The corresponding sum 
\begin{equation}
Q_N(r,0) \ = \  \sum_{\lambda \vdash r} \ d^2(\lambda)
\label{un_integral}
\end{equation}
and its many generalizations have wide applications in different branches of mathematics and have been studied 
for years, see Refs.\cite{regev81,regev15,novak11} (and references therein). We are not aware if a sum of the form 
(\ref{QSUN_A}) have been studied before except for the case $q=1$ \cite{green82}. 
Therefore, in this Appendix we evaluate $Q_N(r,q)$ for all $N$, $q$ and various values of $r$. This enables us to compute 
the large $N$ behaviour of functions $Q_N(r,q)$ and $H_N(h,q)$. 
Also, we derive an asymptotic expansion of the $SU(N)$ integral valid for large values of $h$.

\subsection{Values of $Q_N(r,q)$ for fixed $r$}

Without loss of generality we suppose that $q\geq 0$ in what follows.
First, let us remind that for $q=0$ there is well-known result 
\begin{equation}
Q_N(r,0) \ = \  r! \ , \ r\leq N \ .
\label{QUN}
\end{equation}
Second, it is easy to calculate the value for $r=0$. One finds 
\begin{equation}
Q_N(0,q) \ = \  (qN)! \ A_N(q) \ . 
\label{QSUN_r0}
\end{equation}
The function $A_N(q)$ appears in the solution for the integral (\ref{Z_def_sun}) and equals
\begin{equation}
A_N(q) \ = \ \frac{G(N+1)G(q+1)}{G(N+q+1)} \ ,
\label{ANq}
\end{equation}
where $G(k)$ is the Barnes function.
Third, we note that the convenient representation for $Q_N(r,q)$ can be deduced from (\ref{QSUN_A}) and reads 
\begin{equation}
Q_N(r,q) \ = \  (r+qN)! \ A_N(q) \ B_N(r,q) \ ,
\label{QSUN_small}
\end{equation}
while exact values of $B_N(r,q)$ one calculates either from (\ref{QSUN_A}) or from the following relation 
\begin{equation}
B_N(r,q) \ = \ \sum_{k_1+\cdots +k_N=r} \ \frac{r!}{k_1! \cdots k_N!} \ \frac{A_N(k_i+q)}{A_N(q)} \ ,
\label{BNrs}
\end{equation}
\begin{equation}
A_N(k_i+q) \ = \ \mbox{det} \ \frac{1}{( k_i+i-j+q )!} \ .
\label{ANqdet}
\end{equation}
We had computed $B_N(r,q)$ by "brute force" method for values $r\in [0,11]$ by summing over all partitions in (\ref{QSUN_A}). 
To reveal the structure of these coefficients we give here first three values 
\begin{equation}
B_N(0,q) \ = \ B_N(r,0) \ = \  1  \ ,
\label{BSUN_r0}
\end{equation}
\begin{equation}
B_N(1,q) \ = \ \frac{1}{y}   \ ,
\label{BSUN_r1}
\end{equation}
\begin{equation}
B_N(2,q) \ = \ \frac{1}{y} \ \frac{N^2 y-1}{N^2 y^2-1}  \ ,
\label{BSUN_r2}
\end{equation}
\begin{equation}
B_N(3,q) \ = \ \frac{N^4 y^2 -N^2(3 y+2)+4}{y(N^2 y^2-1)(N^2 y^2-4)}  \ ,
\label{BSUN_r3}
\end{equation}
where we denoted $y=1+q/N$. These expressions are valid for $r\leq N$. 
We would like also to quote result for $q=1$ 
\begin{equation}
B_N(r,1) \ = \ \frac{r! N!}{(r+N)!} \ L_r^{(N)}(1) \ ,
\label{BSUN_q1}
\end{equation}
where $L_r^{(N)}(x)$ is the generalized Laguerre polynomial.
This formula can be obtained from formulas in \cite{green82} and (\ref{QSUN_small}). 
All our coefficients agree with (\ref{BSUN_q1}) if $q=1$.

The knowledge of first eleven coefficients $B_N(r,q)$ is sufficient to compute the function $H_N(h,|q|)$ in Eq.(\ref{Hfunct1}) 
up to ${\cal O}(h^{24})$. In reality, we {\it conjecture} that the following expression 
\begin{equation}
H_N(h,|q|) \ = \ \exp \left [ N^2 \left ( \frac{h^2}{4} - P(x) \right ) \right ] \ , 
\label{Hfunct_res}
\end{equation}
where $x=q/N$ and 
\begin{equation}
P(x) \ = \  \sum_{m=1}^{\infty} x^{m}(-1)^{m+1} \sum_{k = 0}^{\infty} \left(\frac{h}{2}\right)^{2  k+2}\frac{\Omega(k, m)}{k + 1} \ , 
\label{Px_def}
\end{equation}
\begin{eqnarray}
\Omega(k, m) = \frac{4^k \Gamma[\frac{m}{2} + k +1] \Gamma[m + 2 k]}{
\Gamma[\frac{m}{2}+1] \Gamma[m] \Gamma[k+2] \Gamma[2 k+2]}  
\end{eqnarray}
gives an exact representation for the function $H_N(h,|q|)$ to all orders in $h$ in the large $N$ limit. 
First of all, this formula does reproduce correctly all coefficients in $H_N(h,|q|)$ up to ${\cal O}(h^{24})$. 
Secondly, the corresponding exact large $N$ representation for the function $B_N(r,q)$ can be obtained by 
expanding the right-hand side of Eq.(\ref{Hfunct_res}) into power series in $h$, thus recovering representation 
(\ref{Hfunct1}). Combining the result for $B_N(r,q)$ with (\ref{QSUN_small}) and with asymptotic expansion of the function 
$A_N(q)$, Eq.(\ref{ANq_as}), one obtains exact large $N$ representation for the function $Q_N(r,q)$ 
\begin{equation}
Q_N(r,q) \ = \ \frac{(r+qN)!}{N^{qN}} \ e^{N^2 f(x)} \ B_r(1!c_1,2!c_2\ldots,r!c_r) \ . 
\label{QN_exact_largeN}
\end{equation}
Here, $f(x)$ is given by Eq.(\ref{fx_as}) and $B_r(1!c_1,2!c_2\ldots,r!c_r)$ is the complete exponential Bell polynomial. 
The coefficients $c_i$ are functions of $x=q/N$ that can be derived from $P(x)$ and written in several different but equivalent forms. 
For example, one has $c_1= (1+x)^{-1}$ and for $i>1$
\begin{eqnarray}
\label{coeff_ci}
c_i &=&  - \frac{x}{i N^{2(i-1)}} \ \left [ 
C_{i-1}\ _3F_2\left ( i- \frac{1}{2}, i, i+\frac{1}{2} \ ; \ \frac{1}{2}, \frac{3}{2} \ ; \ x^2 \right)  \right.  \\ 
&-& \left. x 4^{i-1} \  _3F_2\left ( i, i+\frac{1}{2}, i+1 \ ; \ \frac{3}{2},  2 \ ; \ x^2 \right ) \right ] \ ,   \nonumber 
\end{eqnarray}
where $_3F_2$ is the hypergeometric function and $C_i$ are the Catalan numbers. 
We performed an extensive numerical comparison of our conjectured 
formula with Eq.(\ref{QSUN_A}) and found perfect agreement with $N$ increasing. 

In the main text we need small and large $x$ representation of the function $P(x)$. 
The first one can be obtained directly from (\ref{Px_def}) by summing over $k$ for fixed small values of $m$. 
One finds on this way 
\begin{eqnarray}
&&P(x) = x \left ( 1 - \sqrt{1 - h^2} +  \ln \frac{1 + \sqrt{ 1 - h^2}}{2} \right ) + 
\frac{x^2}{4} \ln (1 - h^2) \nonumber  \\ 
&&+ \frac{x^3}{6} \left(\frac{ 1  }{(1 - h^2)^{3/2}} -1 \right)
- \frac{x^4}{24}\left(\frac{1 + 12 \left(\frac{h}{2}\right)^2}{ (1 - h^2)^3} -1 \right)   \nonumber   \\ 
&& + \frac{x^5}{60} \left(\frac{1 + 42 \left(\frac{h}{2}\right)^2 + 96 \left(\frac{h}{2}\right)^4}{(1 - h^2)^{9/2}}-1 \right) 
+ {\cal O}\left ( x^6 \right ) \ .
\label{x_small_series}
\end{eqnarray}
To get convenient large $x$ representation we first sum over $m$ at fixed $k$ in (\ref{Px_def}) and then re-expand the result 
obtained at large $x$. This yields after summation over $k$
\begin{eqnarray}
&&P(x) = \frac{h^2}{4}  + x  \left[ 1 - \sqrt{1 + \frac{ h^2}{x^2}} +  
\ln \frac{1 + \sqrt{ 1 +\frac{h^2}{x^2}}}{2}  \right] +  \frac{1}{4} \ln \left ( 1 + \frac{h^2}{x^2} \right )  + \nonumber  \\
&& +  \frac{1}{6 x} \left(\frac{ 1  }{(1 + \frac{h^2}{x^2})^{3/2}} -1 \right) - 
\frac{1}{24x^2}\left(\frac{1 - 12 \left(\frac{h}{2x}\right)^2}{ (1 + \frac{h^2}{x^2})^3} -1 \right) + \nonumber  \\
&&  + \frac{1}{60x^3} \left(\frac{1 - 42 \left(\frac{h}{2x}\right)^2 + 
96 \left(\frac{h}{2x}\right)^4}{(1 + \frac{h^2}{x^2})^{9/2}}-1 \right)  + {\cal O}\left ( x^{-4} \right ) \ .
\label{x_large_series}
\end{eqnarray}
Finally, we need large $N$ expansion for the function $A_N(q)$ from (\ref{ANq}). It can be calculated by making use 
of the well-known asymptotic expansion for the Barnes function. One gets for the leading term
\begin{equation}
N^{q N} \ A_N(q) \ = \ e^{N^2 f(x) + {\cal{O}}(N)} \ , \ x=q/N \ ,
\label{ANq_as}
\end{equation}
\begin{equation}
f(x) \ = \ \frac{3}{2} \ x - \frac{1}{2} \ (1+x)^2\ln (1+x) + \frac{1}{2} \ x^2 \ln x  \ .
\label{fx_as}
\end{equation}

\subsection{Asymptotic expansion for large $h$}

Some expansions of the function $Q_N(r,q)$ can be derived from the equality
\begin{equation}
\sum_{r=0}^{\infty}\frac{ (\frac{z}{2})^{2r+|q|N}}{r!(r+|q|N)!} \ Q_N(r,q)  \ = \
{\rm \det} I_{i-j+q} (z) 
\label{QN_det}
\end{equation}
which follows from the calculation of the integral
\begin{equation}
Z_N(z)  \ = \ \int_{SU(N)}  dU  \ e^{ z {\rm {Re  Tr}}U  } \ = \ \sum_{q=-\infty}^{\infty} \ \det I_{i-j+q} (z) \ .
\label{sun_integral}
\end{equation}
Expanding the integrand in the Taylor series and computing the group integrals with the help of (\ref{Q_def_app}) leads 
to the left-hand side of the Eq.(\ref{QN_det}) summed up over $q$, while the right-hand side can be obtained by the direct 
calculation through an explicit parametrization of the group traces and the invariant measure.
When $z$ is large the main contribution to the sum comes from terms away from $r=0$, the larger $z$ the larger terms in the sum 
are important. In practice it is somewhat simpler to compute $Z_N(z)$ by the saddle-point method when $z$ is large. 
Parametrizing $SU(N)$ matrices through its eigenvalues one obtains
\begin{equation}
Z_N(z)  \ = \ \sum_{q=-\infty}^{\infty} \frac{1}{N!}
\int_0^{2\pi}\prod_{i=1}^N \frac{d\omega_i}{2\pi} \ \prod_{i<j} 4 \sin^2\left ( \frac{\omega_i - \omega_j}{2} \right )
\ e^{ z \sum_{i=1}^N\cos\omega_i + iq\sum_{i=1}^N \omega_i }  \ . 
\label{sun_integral_param}
\end{equation}
Expanding the integrand around non-trivial saddle-point $\omega_k=i\arcsinh\frac{q}{z}$ yields 
\begin{equation}
Z_N(z) = \sum_{q=-\infty}^{\infty} \frac{e^{N \sqrt{z^2+q^2} - N q \arcsinh\frac{q}{z}}}{(2 \pi)^N N!}
\int_{\mathbb{R}^N} d^N \omega 
\prod_{i<j} ( \omega_i - \omega_j )^2
e^{ -\frac{1}{2} \sqrt{z^2+q^2} \sum_{i=1}^N \omega_i^2}  \ . 
\label{sun_integral_selberg}
\end{equation}
The integral obtained is of the Selberg type which can be found , {\it e.g.} in the review \cite{selberg_integral}. 
Comparing the result of the integration with Eq.(\ref{sun_integral}) and taking $z=N h$, $q = N x$ one finds 
\begin{eqnarray}
&&\det I_{i-j+q} (N h) \ \simeq \ \frac{1}{(2 \pi)^{\frac{N}{2}}} \ 
\frac{G(N+1)}{N^{\frac{1}{2}N^2}}  \nonumber \\
&&\exp \left \{ N^2 \left [ \sqrt{h^2 + x^2} - \frac{1}{4} \ \ln (h^2 + x^2) - 
x \ln \left ( \frac{x}{h} + \sqrt{1+\frac{x^2}{h^2}}  \right ) \right ]  \right \} \ .
\label{detsun_as}
\end{eqnarray}
Substituting this expansion on the right-hand side of (\ref{QN_det}) one can derive the corresponding
asymptotics for the function $Q_N(r,q)$ at large $r$. 

Let us finally mention that in the $U(N)$ case $q=x=0$ and the formula (\ref{detsun_as}) reproduces the GWW solution 
in the weak-coupling (large $h$) region. Indeed, using the asymptotic expansion for the Barnes function one can easily obtain 
\begin{equation}
\lim_{N \rightarrow \infty} \frac{1}{N^2} \ln \det I_{i-j} (N h)
\ = \ h - \frac{1}{2} \ \ln h - \frac{3}{4} \ .
\label{GWW_large_h}
\end{equation}

\end{document}